\documentclass[twocolumn,showpacs,fleqn,nobibnotes]{revtex4}

\usepackage{amsmath}
\usepackage{graphicx}
\usepackage{float}
\usepackage{subfigure}

\def\lsim{\raise0.3ex\hbox{$<$\kern-0.75em\raise-1.1ex\hbox{$\sim$}}}
\def\gsim{\raise0.3ex\hbox{$>$\kern-0.75em\raise-1.1ex\hbox{$\sim$}}}

\begin{document}
\newcommand\ie {{\it i.e.}}
\newcommand\eg {{\it e.g.}}
\newcommand\etc{{\it etc.}}
\newcommand\cf {{\it cf.}}
\newcommand\etal {{\it et al.}}
\newcommand{\be}{\begin{eqnarray}}
\newcommand{\ee}{\end{eqnarray}}
\newcommand{\jp}{$ J/ \psi $}
\newcommand{\pp}{$ \psi^{ \prime} $}
\newcommand{\ppp}{$ \psi^{ \prime \prime } $}
\newcommand{\dd}[2]{$ #1 \overline #2 $}
\newcommand\noi {\noindent}

\title{Quarkonium plus prompt-photon associated hadroproduction and nuclear shadowing}
\author{C. Brenner Mariotto$^{a}$ and M.V.T. Machado$^{b}$}

\affiliation{$^a$ Instituto de Matem\'atica, Estat\'{\i}stica e F\'{\i}sica, Universidade Federal do Rio Grande\\
Caixa Postal 474, CEP 96201-900, Rio Grande, RS, Brazil\\
$^b$ Centro de Ci\^encias Exatas e Tecnol\'ogicas, Universidade Federal do Pampa
Campus de Bag\'e, Rua Carlos Barbosa. CEP 96400-970. Bag\'e, RS, Brazil
}

\begin{abstract}
The quarkonium hadroproduction in association with a photon at high energies provides a probe of the dynamics of the strong interactions as it is dependent on the nuclear gluon distribution. Therefore, it could be used to constrain the  behavior of the nuclear gluon distribution in proton-nucleus and nucleus-nucleus collisions. Such  processes are useful to single out the magnitude of the shadowing/antishadowing effects in the nuclear parton densities. In this work we investigate the influence of nuclear effects in the production of $J/\psi + \gamma$ and $\Upsilon + \gamma$ and estimate the transverse momentum dependence of the nuclear modification factors. The theoretical framework considered in the $J/\psi$ ($\Upsilon$) production associated with a direct  photon at the hadron collider is the non-relativistic QCD (NRQCD) factorization formalism.

\end{abstract}

\maketitle
\section{Introduction}
In recent years, quarkonium hadroproduction has become the subject of intense theoretical and experimental investigation. The main reason is that production and decays of heavy quarkonia have been an ideal laboratory to investigate Quantum Chromodynamics (QCD). Their large masses provide a hard scale which allows us to use perturbative QCD techniques. Basically, there exist three distinct formalisms for quarkonium production. The simplest one is the color evaporation model (CEM) \cite{CEM}, where the hadronization of the $Q\bar{Q}$ pairs into quarkonia is assumed to be dominated by long-distance fluctuations of gluon fields which motivates a statistical treatment of color. In the color-singlet model (CSM) \cite{CSM}, quarkonium is viewed as a non-relativistic color-singlet bound state of a $Q\bar{Q}$ pair with definite angular momentum quantum numbers. Production rates for a quarkonium state are computed by calculating the production of a heavy quark pair which is constrained  to be in a color-singlet state and have the same angular momentum quantum numbers as the physical quarkonium. In the non-relativistic QCD (NRQCD) factorization formalism \cite{NRQCD}, non-perturbative aspects of quarkonium production are organized in an expansion in powers of $\nu$, the relative velocity of the $Q\bar{Q}$ in quarkonia. For production of $S$-wave quarkonia, the results of the CSM are recovered in the limit of $\nu \rightarrow 0$. In addition, in the NRQCD formalism new quarkonium production mechanisms are now possible since it is no longer required that the $Q\bar{Q}$ produced in the short-distance process have the same color and angular momentum quantum numbers as the quarkonium state. Recently, several computations of the QCD corrections to the inclusive quarkonium hadroproduction processes have shed some light on the robustness and/or deficiencies of those models. Comprehensive reviews of recent developments in the theory of quarkonium production can be found in Refs. \cite{Lansberg,Kramer:2001}.

Beside the studies of inclusive production, efforts are being made to obtain improved theoretical predictions for complementary observables to the inclusive yield, such the hadroproduction of $J/\psi$ and $\Upsilon$ in association with a photon. In the framework of CSM, the associated production of  $J/\psi+\gamma$ at a hadron collider was first proposed as a good channel to investigate the gluon distribution in the proton with a relatively clean signal \cite{Kim1}. In Ref. \cite{Roy}, such a process at the Tevatron energy has been considered in the CSM at LO, and the results show that the contribution from the gluon fusion sub-process is dominant over that from the fragmentation process (the same occurs at  the LHC \cite{Basu}). The color-octet contributions were investigated in Ref. \cite{Kim2} and it was found they are dominant in the large $p_T$ region. Recently, in Ref. \cite{Wang} the effect of the NLO QCD corrections to  $J/\psi\,(\Upsilon)+\gamma$ hadroproduction at the LHC has been investigated. In Ref. \cite{Lansberg2} the real next-to-next-to-leading (NNLO) order QCD contribution to hadroproduction of a $J/\psi\,(\Upsilon) +\gamma$ via color singlet transitions for the inclusive case has been addressed. In this Letter, we examine the production of associated $J/\psi+\gamma$ and $\Upsilon +\gamma$ at large $p_T$ within the NRQCD approach in proton-nucleus and nucleus-nucleus collisions. Such processes are relatively clean because the produced large $p_T$ quarkonium is easy to detect through its leptonic decay modes and the quarkonium
's large $p_T$ is balanced by the associated high energy photon. At the LHC energy, the leading contributions are dominated by gluon induced hard processes and then the quarkonium production associated with a direct photon will be strongly dependent on the nuclear gluon distribution.

Our goal is to use the $J/\psi+\gamma$ and $\Upsilon +\gamma$ processes as auxiliary observables to constrain the nuclear gluon distribution. This is motivated by similar investigations on inclusive heavy quark, quarkonium and prompt photon production in central proton-nucleus and nucleus-nucleus collisions (See e.g. Refs. 
\cite{vogt1,vogt2,vogt3,vogt4,vic_luiz1,ABMG:2006,BMG:2008}). 
One of the nuclear effects which is expected to modify the behavior of gluon distribution is the nuclear shadowing. This effect has been observed in the nuclear structure functions by different experimental collaborations  \cite{arneodo,e665} in the study of the deep inelastic lepton scattering (DIS) off nuclei. The modifications on $F_2^A (x,Q^2)$ depend on the parton momentum fraction $x$. While for momentum fractions $x < 0.1$ (shadowing region)
and $0.3 < x < 0.7$ (EMC region), a depletion is observed in the nuclear structure functions, in the intermediate region ($0.1 < x < 0.3$) it is verified an enhancement known as antishadowing. These experimental results strongly constrain the behavior of the nuclear quark distributions, whereas the the  nuclear gluon distribution is still an open question due to the scarce experimental data in the small-$x$ region and/or for observables strongly dependent on the nuclear gluon distribution.

This Letter is organized as follows. In next section, we summarize the main formulas concerning the process $J/\psi+\gamma$ in the NRQCD formalism and define the nuclear modification factors  for proton-nucleus and nucleus-nucleus collisions, $R_{pA}$ and $R_{AA}$, respectively. In last section we present the numerical results considering the more recent nuclear parton parameterizations and estimating the transverse momentum dependence of the nuclear modification factors at the LHC energies.

\section{Quarkonium production associated with a direct photon}
 As long as $J/\psi+\gamma$ is produced at small longitudinal momentum fraction, $x_F\ll 1$, the gluon fusion channel dominates over the $q\bar{q}$ annihilation process.  Therefore, at high energies and at leading order (LO), the process $g+g\rightarrow J/\psi + \gamma$ contributes at the partonic level with six Feynman diagrams, which is similar to that of $g+g\rightarrow J/\psi + g$  in inclusive $J/\psi$ hadroproduction. The signal we focus on is the production of a $J/\psi$ and an isolated photon produced back-to-back, with their transverse momenta balanced.
The LO cross section is obtained by convoluting the partonic cross section with the parton distribution function (PDF), $g(x,\mu_F)$, in the proton, where $\mu_F$ is the factorization scale. At NLO expansion on $\alpha_s$, there are one virtual correction and three real corrections processes, as shown in Ref. \cite{Wang}. In the NRQCD formalism, the contributing subprocesses are $q+\bar{q}\rightarrow \, ^{2S+1}\!L_J+\gamma$  and $g+g\rightarrow\,^{2S+1}\!L_J+\gamma$. The Fock-components that contribute to $J/\psi$ production are the color-singlet $^3\!S_1^{[1]}$ state and the color-octet states $^3\!S_1^{[8]}$, $^1\!S_0^{[8]}$ and $^3\!P_{0,1,2}^{[8]}$. The color-singlet $^3\!S_1$ state contributes at ${\cal O} (1)$ but the color-octet channels all contribute higher orders in $\nu$ (the relative velocity between the heavy quarks).

We are interested here in the quarkonium production in a nuclear medium. In order to get the $J/\psi+\gamma$ yield in $pA$ and $AA$ collisions, a shadowing-correction factor has to be applied to the $J/\psi$ yield obtained from the simple superposition of the equivalent number of $pp$ collisions. This shadowing factor can be expressed in terms of the ratios $R_i^A$ of the nuclear parton distribution functions in a nucleon of a nucleus $A$ to the PDF in the free nucleon. Most of shadowing models provide the nuclear ratios at a given value of $Q_0^2$  and then evoluted through the DGLAP evolution equations \cite{dglap} to LO accuracy. Only very recently, the nuclear PDFs have been available at NLO accuracy. Therefore, in what follows we will consider a LO calculation for the nuclear modification factors in order to be consistent with the limitation of shadowing models. In this respect our analysis cannot be considered as fully NLO and should be updated once the NLO calculation in NRQCD approach is available for the nucleon case. On the other hand, this is not an important limitation to our results as we will discuss later on.

In order to obtain the transverse momentum ($p_T$) distribution for the process $g+g\rightarrow J/\psi + \gamma$, we express the differential cross section as
\begin{eqnarray}
\frac{d^2\sigma}{dydp_T}=\int dx_1g_A(x_1,\mu_F^2) g_B(x_2,\mu_F^2)\frac{4x_1x_2p_T}{2x_1-\bar{x}_Te^y}\frac{d\hat{\sigma}}{d\hat{t}},\,\,
\label{diffcs}
\end{eqnarray}
where we have defined $\bar{x}_T=2m_T/\sqrt{s}$, with $\sqrt{s}$ being the center of mass energy of the $AB$ system and $m_T=\sqrt{p_T^2+m_{\psi}^2}$ being the transverse mass of outgoing $J/\psi$. The gluon distribution, $g_{A/B}(x,Q^2)$, in the hadron A/B is evaluated at factorization scale $\mu_F$. The common transverse momentum of the outgoing particles is $p_T$ and $y$ is the rapidity of outgoing $J/\psi$ having mass $m_{\psi}$. The variables $x_1$ and $x_2$ are the momentum fractions of the partons, where $M^2/s\leq x_1<1$ ($M$ is the invariant mass of $J/\psi+\gamma$ system) and $x_2$ can be written in terms of other variables as
\begin{eqnarray}
x_2=\frac{x_1\bar{x}_Te^{-y}-2\tau}{2x_1-\bar{x}_Te^y}, \,\,\,\,\mathrm{with} \,\,\,\,\tau=\frac{m_{\psi}^2}{s}.
\end{eqnarray}

In the NRQCD formalism, the cross section for the production of a quarkonium state $H$ is written as $\sigma (H)=\sum_n c_n\langle 0|O_n^H|0 \rangle$, where the short-distance coefficients $c_n$ are computable in perturbation theory. The $\langle 0|O_n^H|0 \rangle$ are matrix elements of NRQCD operators of the form
\begin{equation}\label{Matrix}
\langle 0|O_n^H|0 \rangle = \sum_X \sum_{\lambda} \langle 0|\kappa_n^{\dagger}|H(\lambda) + X\rangle \langle H(\lambda) + X |\kappa_n|0 \rangle.
\end{equation}
The $\kappa_n$ is a bilinear in heavy quark fields
which creates a $Q \overline{Q}$ pair in a state with definite color and angular momentum quantum numbers. Hereafter, we will use a shorthand notation in which the matrix elements are given as $\langle O^H_{(1,8)}(^{2S+1}L_J) \rangle$. The angular momentum quantum numbers of the $Q\overline{Q}$
produced in the short-distance process are given in standard spectroscopic notation, and the subscript refers to the color configuration of the $Q\overline{Q}$: $1$ for a color singlet and $8$ for a color octet.

\begin{figure*}[t]
\includegraphics[scale=0.4]{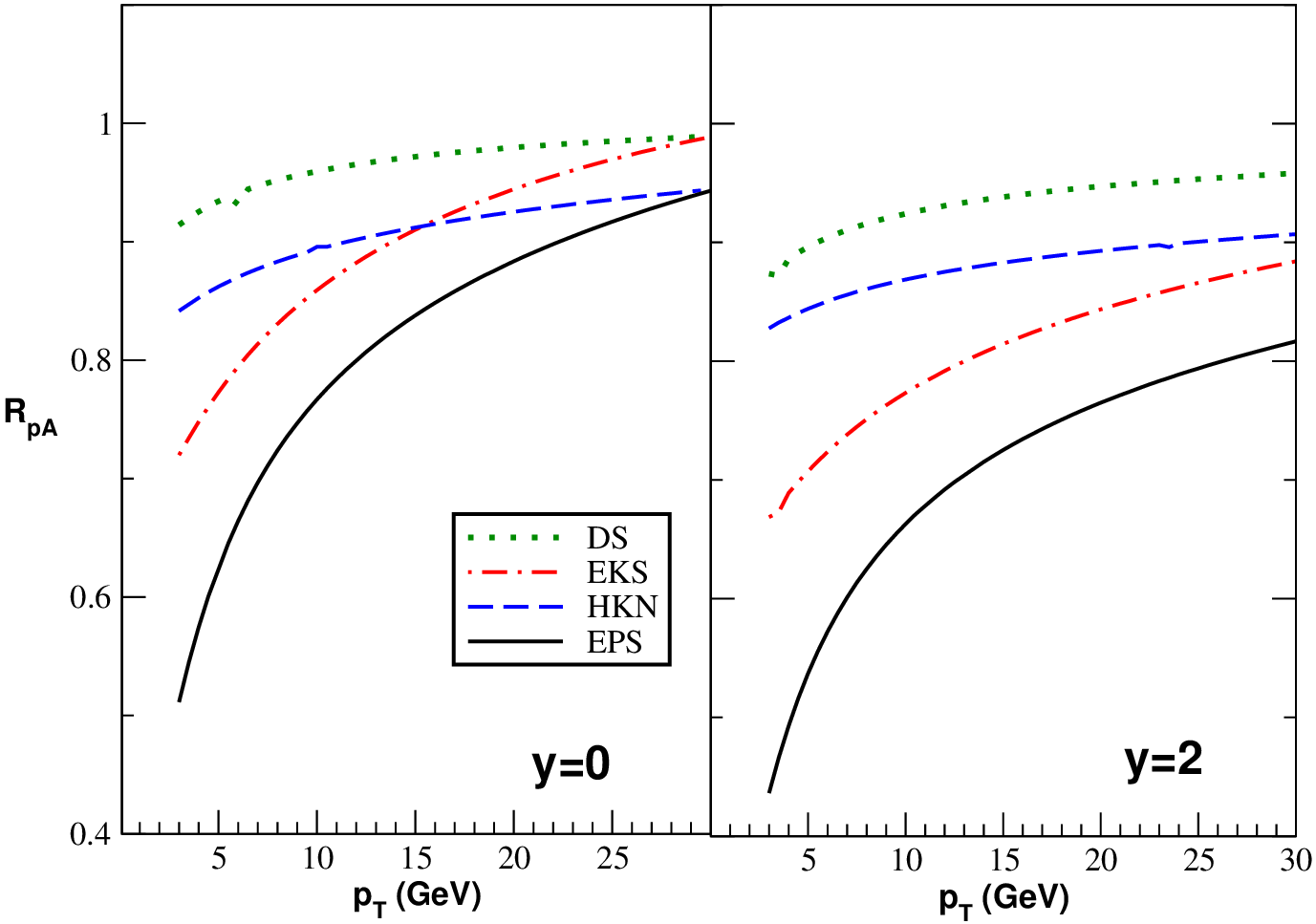} 
\caption{Transverse momentum dependence of the nuclear modification factor $R_{pA}$
in $J/\psi+\gamma$ production in central and forward rapidities at the LHC ($\sqrt{s}=8.8$ TeV), for distinct nPDFs.
} \label{fig:RpA}
\end{figure*}

The parton level differential cross sections relevant for hadroproduction
of $J/\psi + \gamma$, including both color-singlet and color-octet contributions are given below \cite{Mehen,Kim2}:
\begin{eqnarray}\label{GGPsiGa}
{d\hat{\sigma}_{\mathrm{sing}} \over d\hat{t}} & = & \sigma_0 \left[ {16\over 27} \left( {\hat{s}^2 s_1^2 + \hat{t}^2 t_1^2 + \hat{u}^2 u_1^2 \over s_1^2 t_1^2 u_1^2} \right) \langle O_1^{J/\psi}(^3S_1)\rangle \right] ,\nonumber \\
{d\hat{\sigma}_{\mathrm{oct}} \over d\hat{t}} & = &  \sigma_0 \left[ {10\over 9} \left( {\hat{s}^2 s_1^2 + \hat{t}^2 t_1^2 + \hat{u}^2 u_1^2 \over s_1^2 t_1^2 u_1^2} \right)
\langle O_8^{J/\psi}(^3S_1)\rangle \right. \nonumber \\
& + & {6 \over \hat{s} s_1^2 m_c^2} \left(2 \hat{s} + {3 \hat{t} \hat{u} \over 4m_c^2} - {4 \hat{t} \hat{u} \over s_1} \right) \langle O_8^{J/\psi}(^1P_0)\rangle \nonumber \\
 & + & \left. {3\over 2} {\hat{t} \hat{u} \over \hat{s} s_1^2 m_c^2} \langle O_8^{J/\psi}(^1S_0)\rangle \right] ,
\label{Jpsixsection}\end{eqnarray}

In Eqs. (\ref{GGPsiGa}) we have defined $\sigma_0=\pi^2 e_c^2 \alpha \alpha_s^2 m_c/ \hat{s}^2$ (with charm quark mass $m_c=1.5$ GeV) and $s_1 = \hat{s} - 4 m_c^2$, $t_1 = \hat{t} - 4 m_c^2$, and  $u_1 = \hat{u} - 4 m_c^2$. In these formulae, $\hat{s}$, $\hat{t}$, and $\hat{u}$ are the Mandelstam variables, which can be written as
\begin{eqnarray}
\hat{s}=x_1x_2s,\hspace{0.1cm} \hat{t}=m_{\psi}^2-x_2\sqrt{s}m_{T}e^y, \hspace{0.1cm} \hat{u}=m_{\psi}^2-x_1\sqrt{s}m_{T}e^{-y}. \nonumber
\end{eqnarray}

In our numerical calculations the one loop expression for the running coupling, $\alpha_s(\mu_R)$, with $\Lambda_{QCD}=0.2$ GeV and $n_f=4$ is considered. The (renormalization and factorization) scale for the strong coupling and for the evaluation of PDFs is $\mu_F^2=\mu_R^2=(p_T^2+m_{\psi}^2)$, where $m_{\psi}$ is the $J/\psi$ 
mass. 
For numerical values of the NRQCD matrix elements we have used those from Ref. \cite{Matrix}, which are (units of $GeV^3$): $\langle O_1^{J/\psi}(^3S_1)\rangle=1.16$, $\langle O_8^{J/\psi}(^3S_1)\rangle=1.19\times 10^{-2}$, $\langle O_8^{J/\psi}(^1S_0)\rangle=\langle O_8^{J/\psi}(^1P_0)\rangle/m_c^2=0.01$. We have checked that using another set of color octet matrix elements, taken from \cite{Kramer:2001}, our results do not change considerably, since we only calculate the ratio between the production cross sections, and the dependency on the color octet matrix elements cancels out in those ratios. 
In what follows we estimate the differential cross sections for central ($y=0$) and forward ($y=2$) rapidities for the $pp$, $pA$ and $AA$ collisions in order to compute the nuclear modification factors.  From Eq. (\ref{diffcs}), it implies that  the  differential cross section at small values of the $J/\psi$ transverse momentum in  $pPb$ (and  $PbPb$) collisions at the LHC is determined by the behavior of the nuclear gluon distribution at $x_2 \gtrsim 10^{-4} $. It is important to emphasize that smaller values of $x$ contribute at $J/\psi+\gamma$ production in the forward rapidity region which can be measured, e.g., with the CMS and ALICE experiments at LHC.

\begin{figure*}[t]
\includegraphics[scale=0.4]{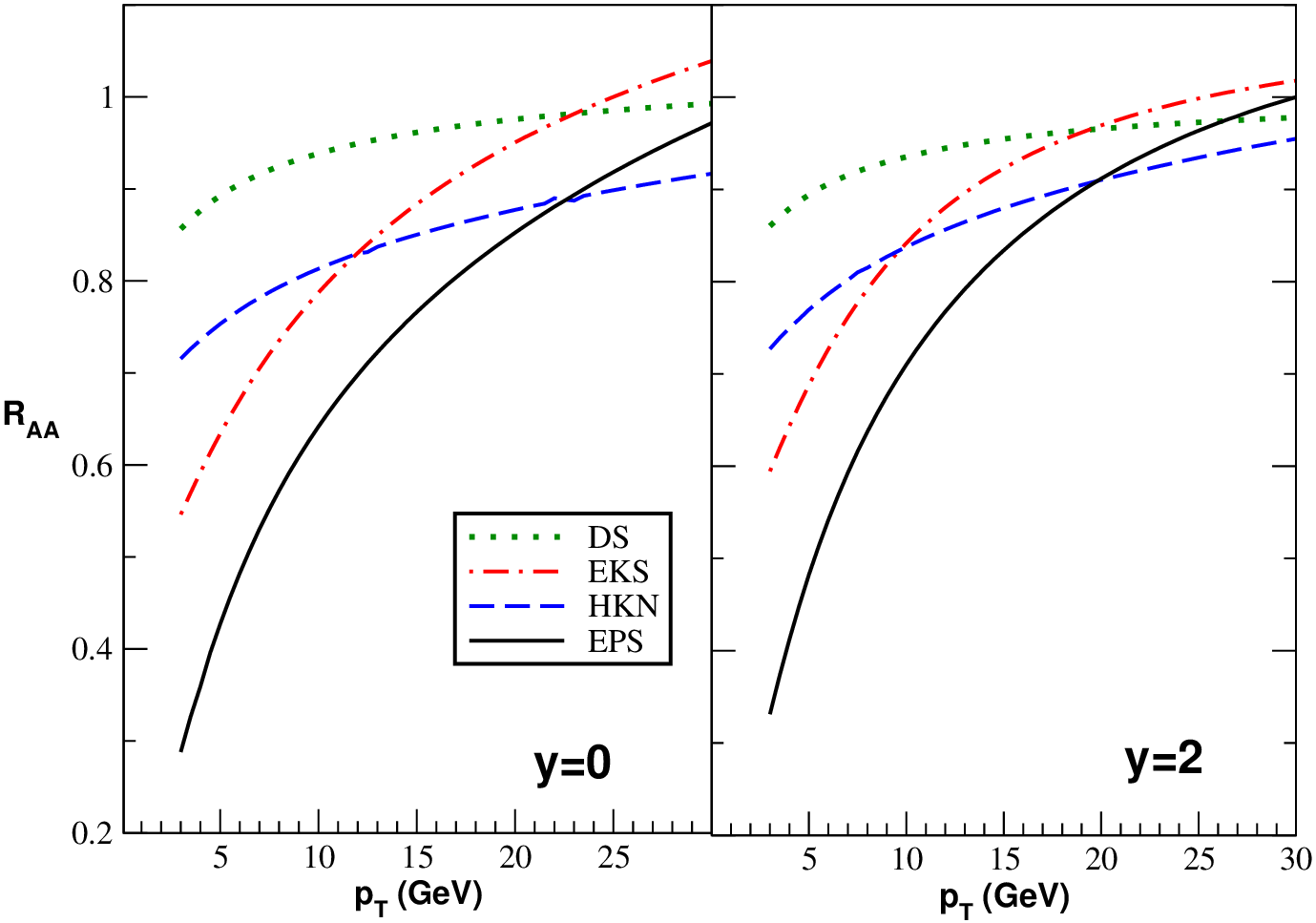} 
\caption{Transverse momentum dependence of the nuclear modification factor $R_{AA}$
in $J/\psi+\gamma$ production in central and forward rapidities at the LHC ($\sqrt{s}=5.5$ TeV), for distinct nPDFs. }
\label{fig:RAA}
\end{figure*}

The main input in the  calculations of $J/\psi+\gamma$ cross sections is the nuclear parton distribution function (nPDF). Over last years several groups has proposed parameterizations for the nPDFs \cite{EKS98,sassot,HKM,HKN,EPS08}, which are based on different assumptions and techniques to perform a global fit of different sets of data using the DGLAP evolution equations. These parameterizations predict very distinct magnitudes for the nuclear effects. For larger values of $x$, the EKS and the EPS nPDFs show antishadowing, while this effect is absent for the HKN and DS parameterizations in the $x \le 10^{-1}$ domain.  While the nuclear shadowing is moderate for DS and HKN parameterizations and somewhat bigger for EKS one, the EPS prediction has a much stronger suppression compared with the other parameterizations. For smaller $x$ around $x\simeq {10}^{-5}$, while DS and HKN parameterizations have about $20\%$ suppression and EKS one have about $40\%$ suppression, for the EPS parameterization this effect goes to almost $80\%$ suppression in the nuclear gluon compared with the $A$ scaled gluon content in the proton.

In what follows we calculate the quarkonium production in association with a direct photon in $pA$ and $AA$ collisions, considering
the nuclear parton distributions discussed above. We then estimate the nuclear modification factor for these predictions. These factors  could be measured at the LHC and we analyze the transverse momentum dependence of them. Since the full NLO calculation of $J/\Psi (\Upsilon)+\gamma$ including the COM contributions is not available yet, and for the sake of simplicity, 
the calculation of the cross sections is here done at LO accuracy. In order of to be consistent, we consider only the LO version of the nPDFs employed (besides, the EKS nPDF is only evolved to leading order). As we are calculating ratios between cross sections, common uncertainties on the normalization of the  $pA$, $AA$ and $pp$ cross sections, e.g. due to higher order contributions, are expected to cancel out in the nuclear modification factors.

In Fig. \ref{fig:RpA} we present our estimates for the transverse momentum dependence of the ratio $R_{pA}$ for central and forward rapidities, defined by
\begin{eqnarray}
R_{pA}  \equiv \left. \frac{d\sigma (pA)}{dy d^2 p_T}\right|_{y=\,0,\,2} / \left.  A \frac{d\sigma
(pp)}{dy d^2 p_T}\right|_{y=\,0,\,2}  \,\,,
\end{eqnarray}
in $pA$ collisions at the LHC ($\sqrt{s}=8.8$ TeV). The results show distinct behaviors of $R_{pA}$ for different nPDF's. The  difference  between central and forward rapidities for $R_{pA}$ comes from the kinematical $x_2$ range probed in the two cases. While for central rapidities the values of $x_2$ are ever larger than $10^{-3}$, in the forward case the minimum value could be  $10^{-4}$, increasing with $p_T$. Consequently, the effects in the nuclear gluon distribution which contribute for the quarkonium production are different in the two rapidities. In the $p_T$ interval shown in Fig. \ref{fig:RpA}, antishadowing effect is not observed as $x_2<0.1$. We have that the nuclear factor is substantially suppressed in the EPS (EKS) case, going down an $0.8-0.7$ at $p_T \simeq 10$ GeV, while in the DS (HKN) case it is almost flat and equal to 0.95 (0.85) in the full $p_T$ range. Moreover, differently from the DS and HKN predictions, the EKS and EPS parameterizations lead to a  strong transverse momentum dependence. Consequently, the determination of the magnitude and $p_T$ dependence of the this nuclear modification factor at the LHC could be useful to determine the properties of the shadowing in the gluon distribution.

The production of $J/\psi+\gamma$ can also be studied in the collision of two heavy nuclei. In what follows  we present our estimates for the transverse momentum dependence of the nuclear modification factor $R_{AA}$  for central and forward rapidities defined by
\begin{eqnarray}
R_{AA}  \equiv \left. \frac{d\sigma (AA)}{dy d^2 p_T}\right|_{y=\,0,\,2}  / \left. A^2 \frac{d\sigma
(pp)}{dy d^2 p_T}\right|_{y=\,0,\,2} \,\,,
\end{eqnarray}
in $AA$ collisions at the LHC ($\sqrt{s}=5.5$ TeV). In this case the nuclear effects are amplified by the presence of two nuclei in the initial state of the collision. Similarly to the $pA$ case, we can see in Fig. \ref{fig:RAA} that the factor $R_{AA}$ is strongly modified by the shadowing effects. For low $p_T$ the suppression is stronger than in the $pA$ case. We can see an anti-shadowing effect appearing in central rapidity for EPS and EKS nPDFs in the region of larger $p_T\simeq 30$ GeV. Our results indicate that $J/\psi+\gamma$ production with $p_T \le 20$ GeV can be used to determine the gluon shadowing effect.

\begin{figure*}[t]
\includegraphics[scale=0.4]{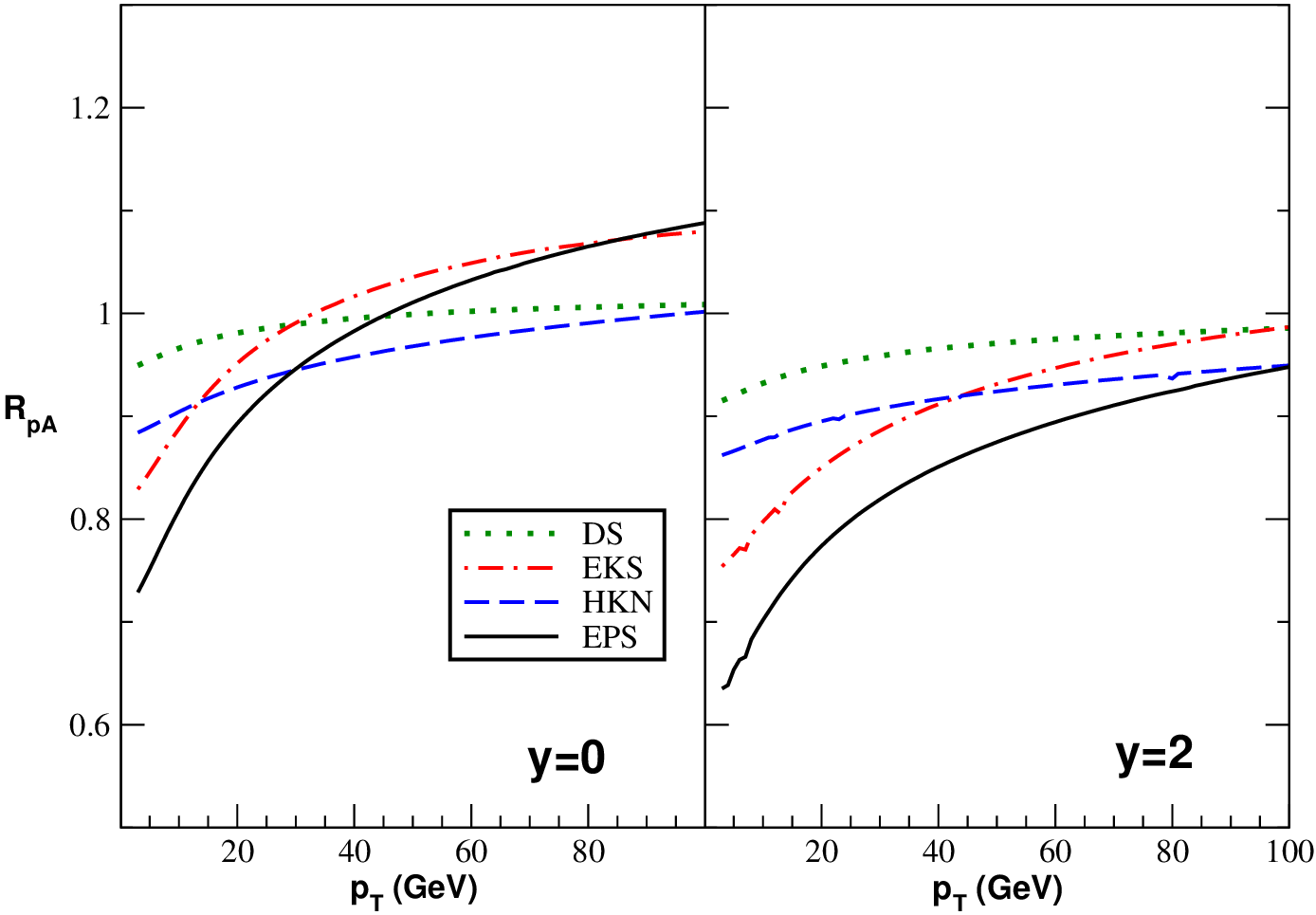} 
\caption{Transverse momentum dependence of the nuclear modification factor $R_{pA}$
in $\Upsilon (1S) +\gamma$ production in central and forward rapidities at the LHC ($\sqrt{s}=8.8$ TeV), for distinct nPDFs. }
\label{fig:RpA_upsi}
\end{figure*}

The production of $\Upsilon(1S)+\gamma$ can be obtained from the expression (\ref{Jpsixsection}) above, by replacing the charm mass and charge by the bottom ones, $m_b=4.7\, GeV$, $e_b$, the $J/\psi$ mass by the $\Upsilon (1S)$ mass, and by using the corresponding color octet matrix elements. We use the values taken from \cite{Braaten:2001}, namely (units of $GeV^3$): $\langle O_1^{\Upsilon}(^3S_1)\rangle=10.9$, $\langle O_8^{\Upsilon}(^3S_1)\rangle=0.02$, $\langle O_8^{\Upsilon}(^1S_0)\rangle=0.136$ and $\langle O_8^{\Upsilon}(^1P_0)\rangle=0$. 
We notice that 
using the alternative set for the color octet matrix elements \cite{Braaten:2001}:  
$\langle O_8^{\Upsilon}(^3S_1)\rangle=0.03$, 
$\langle O_8^{\Upsilon}(^1S_0)\rangle=0$ and 
$5\langle O_8^{\Upsilon}(^1P_0)\rangle/m_b^2=0.139$, 
our results are not significantly changed (there is only a very tiny reduction in $R$ for larger $p_T$). 

In Fig. \ref{fig:RpA_upsi} we present our estimates for the transverse momentum dependence of the ratio $R_{pA}$ for central and forward rapidities,  in $pA$ collisions at the LHC ($\sqrt{s}=8.8$ TeV) producing $\Upsilon (1S)+\gamma$. In this case the $p_T$ range is extended to $100$ GeV, since the distributions tend to be shifted to larger values of $p_T$ due to the larger $\Upsilon$ mass, which makes larger values of $x$ to be accessed, and suggests the existence of antishadowing. As in the previous cases the results show distinct behaviors of $R_{pA}$ for different nPDF's. For central rapidities, the EKS and EPS show antishadowing for larger $p_T$, while this effect is absent for forward rapidities and for the other distributions. For lower $p_T$, the suppression is different for the different nPDFs, being stronger in the forward case. This effect could then be used to discriminate among the nuclear PDF's.

\begin{figure*}[t]
\includegraphics[scale=0.4]{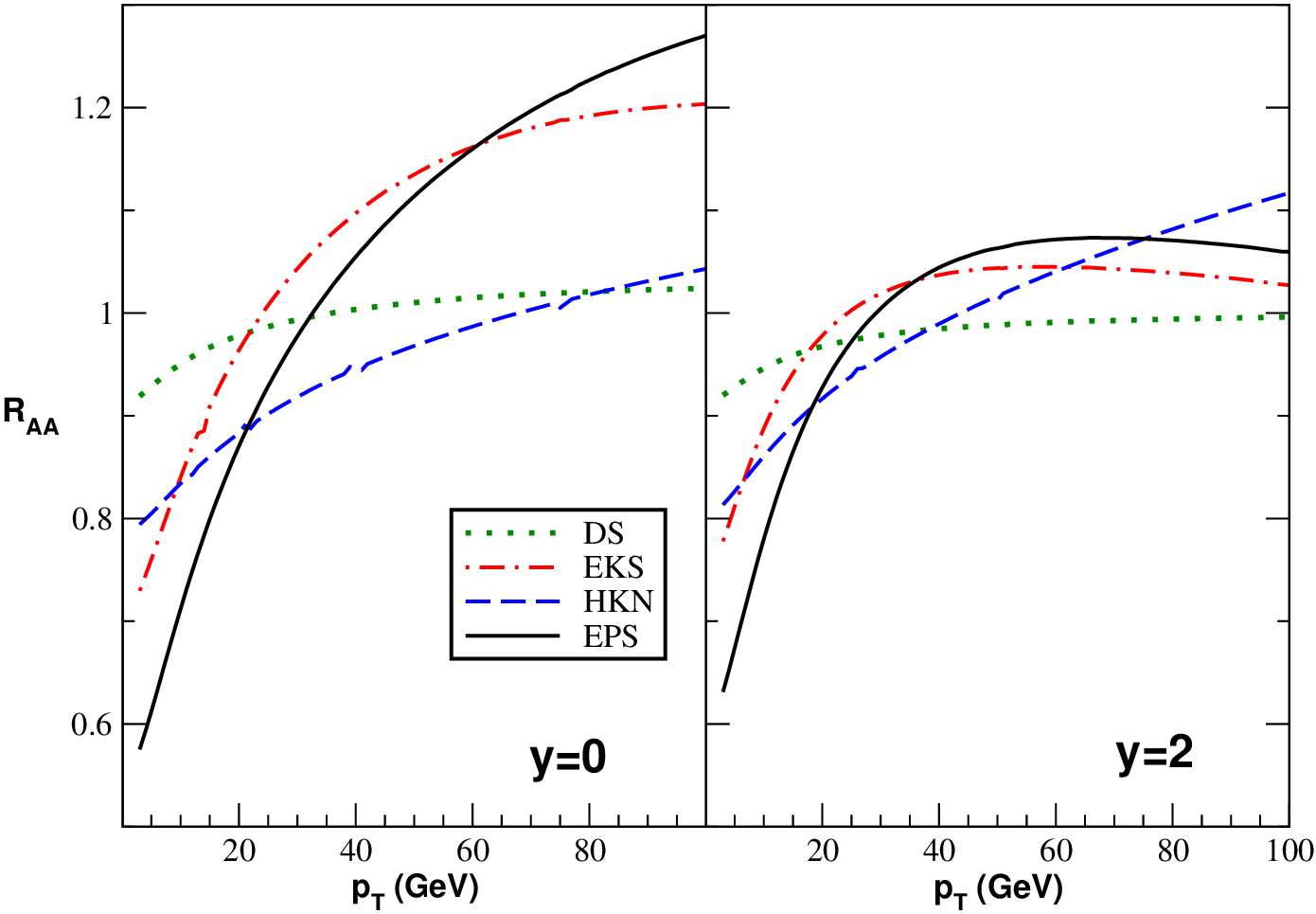}
\caption{Transverse momentum dependence of the nuclear modification factor $R_{AA}$
in $\Upsilon (1S) +\gamma$ production in central and forward rapidities at the LHC ($\sqrt{s}=5.5$ TeV), for distinct nPDFs. }
\label{fig:RAA_upsi}
\end{figure*}

The results for the production of $\Upsilon(1S)+\gamma$ in $AA$ collisions are shown in Fig. \ref{fig:RAA_upsi}, where we present our results for the transverse momentum dependence of the ratio $R_{AA}$ for central and forward rapidities at the LHC ($\sqrt{s}=5.5$ TeV). Here the nuclear effects are more pronounced than in the $pA$ case. The EPS parameterization shows the steepest behavior, presenting shadowing (suppression) for low $p_T$ and antishadowing (enhancement) for larger $p_T$ and central rapidities. The EKS has a similar although less pronounced behavior, and the HKN, which presented an almost flat behavior in the $J/\psi$ case, it now grows steeper. Thus, the HKN versus DS could be discriminated at low $p_T$ and central rapidities, and at high $p_T$ and forward rapidities. In the forward case, antishadowing behavior at high $p_T$ tend to be less important for the EKS and EPS nPDF's as $p_T$ grows. We conclude that shadowing and antishadowing effects could be tested in different $p_T$ and rapidity regions and therefore help in discriminating among the different nuclear PDF's.

Finally, we discuss qualitatively the results and comment on the limitations of present calculations. Let us first concentrate on the $pA$ case where cold matter effects also play an essential role. At leading order, the differential cross section is simply proportional to the product of gluon densities
\begin{eqnarray}
\frac{d\sigma}{dx_1dx_2}(pA\rightarrow J/\psi+\gamma)\propto g_p(x_1,Q^2)g_A(x_2,Q^2)\nonumber\\
\times \delta (x_1x_2s-m_{\psi}^2),\nonumber
\end{eqnarray}
where $x_{1,2}$ are the projectile and target-parton momentum fractions. In this kinematics, the nuclear modification factor reduces to $R_{pA}\simeq R_g^A(x_2)$. In our analysis above we did not take into account the nuclear absorption. In the framework of the probabilistic Glauber model, this effect refers to the probability for the pre-resonant $Q\bar{Q}$ pair to survive to the propagation through the nuclear medium. Therefore, the $J/\psi$ may be sensitive to inelastic rescattering processes in a large nuclei, which spoil the simple relationship between $R_{pA}$ and $R_g^A$. Assuming the factorization between the quarkonium production process and the subsequent possible $J/\psi$ inelastic interaction with nuclear matter, we can write the production cross section as
\begin{eqnarray}
\frac{d\sigma}{dx_2}(pA\rightarrow J/\psi+\gamma)\propto  S_{\mathrm{abs}}(A,\sigma_{J/\psi N}) \nonumber\\ \times \frac{d\sigma_{\mathrm{prod}}}{dx_2}(pA\rightarrow J/\psi+\gamma),\nonumber
\end{eqnarray}
where $S_{\mathrm{abs}}$ denotes the probability for no interaction, or survival probability, of the meson with the nucleus target. In a Glauber model it depends on the $J/\psi-N$ inelastic cross section, $\sigma_{J/\psi N}$, and reads as (for large nucleus):
\begin{eqnarray}
S_{\mathrm{abs}} \simeq \frac{1}{A\sigma_{J/\psi N}}\int d^2b\,\left[1-
\exp\left(-T_A(b)\,\sigma_{J/\psi N}\right) \right],
\end{eqnarray}
with the thickness function $T_A(b)$. As $\sigma_{J/\psi N}$ is not well constrained from data, this is additional uncertainty entering on the nuclear modification factor. In a rough estimation we have $R_{pA}\simeq S_{\mathrm{abs}}(A,\sigma_{J/\psi N})  R_g^A$. The corresponding expression for $AA$ collision can be obtained using similar methods. We quote Ref. \cite{Elena} as an example where the magnitude of these corrections has been studied for the $J/\psi$ production in proton-nucleus and nucleus-nucleus collisions.

Concerning nucleus-nucleus collisions, the situation is more complicated as final state effects can not be disregarded. For instance, the processes of dissociation and recombination of $c\bar{c}$ pairs in the dense medium can be computed through the
co-movers interaction model \cite{Cap05}, which also incorporates also the recombination mechanism
\cite{epjc08}. As mentioned for $pA$ case, nuclear effects in nucleus-nucleus collisions are usually expressed through
the so-called nuclear modification factor, $R^{J/\psi}_{AB}
(b)$, defined as the ratio of the $J/\psi$ yield in $AA$ and
{\it pp} scaled by the
number of binary nucleon-nucleon collisions, $N_{\mathrm{coll}}(b)$. A similar factor can be defined in our case of $J/\Psi + \gamma$ production, having in mind that the prompt photon is insensitive to nuclear matter effects.  We have then for symmetric nuclei,
\begin{eqnarray}
\label{eq:ratioJpsi}
R^{J/\psi}_{AA}(b)&=&
\frac{\mbox{d}N^{J/\psi}_{AA}/\mbox{d}y}{N_{\mathrm{coll}}(b)
  \,\mbox{d}N^{J/\psi}_{pp}/\mbox{d}y} \nonumber \\
&=& \frac{\int\mbox{d}^2s \,
  \sigma_{AA}(b) \, n(b,s) \, P_{\mathrm{sup}}(b,s)
}{\int \mbox{d}^2 s \, \sigma_{AA} (b) \, n(b,s)} \;.
\end{eqnarray}
where $\sigma_{AA}(b) = 1 - \exp [-\sigma_{pp}\, A^2\, T_{AA}(b)]$ and
$T_{AA}(b) = \int\mbox{d}^2s T_A(s)T_A(b-s)$ is the nuclear overlapping
function. One has that  $T_A(b)$ is obtained from Woods-Saxon nuclear densities, and
\begin{eqnarray}
\label{eq:nbin}
P_{\mathrm{sup}}(b,s) & = & S_{J/\psi}^{sh}(b,s) \,
  S^{abs}(b,s) \, S^{co}(b,s), \\
n(b,s) & = & \sigma_{pp} A^2 \, T_A(s)\, T_A(b-s)/\sigma_{AA}(b)\;,
\label{nbin2}
\end{eqnarray}
where the number of
binary nucleon-nucleon collisions at impact parameter $b$ is given by, $N_{\mathrm{coll}}(b)=\int d^2s\,n(s,b)$.
The three factors appearing in Eq.~(\ref{eq:nbin}), $S^{sh}$, $S^{abs}$ and $S^{co}$, denote the
effects of shadowing, nuclear absorption, and interaction with the
co-moving matter, respectively.

As referred above, the nuclear absorption is usually interpreted as suppression of
$J/\psi$ yield because of multiple scattering of a $c\bar{c}$ pair
within the nuclear medium. At low energies the primordial spectrum of particles created in
scattering off a nucleus is mainly altered by interactions
with the nuclear matter they traverse on the way out to the detector
and energy-momentum conservation (it has been shown that it becomes
a minor effect at $AA$ collisions at LHC). For nucleus-nucleus collisions
these effects can be combined into the generalized suppression factor
\begin{eqnarray}
\label{eq:Sabs}
S^{abs} &=& \frac{
        \left[1 - \exp \left(-\beta(x_1) \sigma_{c\bar{c}} AT_A(b) \right) \right]}
       {\beta(x_1)\beta(x_2)\, \sigma_{c\bar{c}}^2 \,A^2 \, T_A(s) T_A(b-s)}
\nonumber \\
&\times& \left[1 - \exp \left(-\beta(x_2) \,\sigma_{c\bar{c}}\, AT_A(b-s)\right) \right]\;,
\end{eqnarray}
where $x_{1,2} = (\sqrt{x_{\rm F}^2 - 4M^2/s} \pm x_{\rm F})/2$, and
$\beta(x_{1,2}) = (1-\epsilon) + \epsilon x_{1,2}^\gamma$ determines
both absorption and energy-momentum conservation. The parameters $\gamma $, $\epsilon $
and $\sigma_{c\bar{c}}$ can be adjusted to describe collider data. Secondly, coherence effects will lead to nuclear shadowing for both soft
and hard processes and therefore for the production of
heavy flavor. Shadowing factor $S_{J/\psi}^{sh}(b,s)$ can be calculated
within the Glauber-Gribov theory, and here it is already included in the nuclear PDF ratios.

The final state effects can be addressed taking, for example, the co-movers interaction model \cite{epjc08}.
Assuming a pure longitudinal expansion and boost invariance of the
system, the rate equation which includes both dissociation and
recombination effects for the density of charmonium at a given
production point at impact parameter $s$ reads
\begin{eqnarray}
 \frac{d N_{J/\psi} (b,s,y)}{d \tau} &=& -\frac{\sigma_{co}}{\tau}
\Big[ N^{co}(b,s,y) \,N_{J/\psi}(b,s,y) \nonumber \\
   & - &  N_c (b,s,y)\, N_{\bar{c}}(b,s,y) \Big] \;,
   \label{eq:rateeq}
\end{eqnarray}
where $N^{co}$, $N_{J/\psi}$ and $N_{c (\bar{c})}$ is the density of
comovers, $J/\psi$ and open charm, respectively, and $\sigma_{co}$ is
the interaction cross section for both dissociation of charmonium with
co-movers and regeneration of $J/\psi$ from $c\bar{c}$ pairs in the system
averaged over the momentum distribution of the participants. It is the
constant of proportionality for both the dissociation and recombination
terms due to detailed balance $ N_{J/\psi}\, N^{co} = N_{c}\,N_{\bar{c}}.$
The solution of Eq.~(\ref{eq:rateeq}) can be approximated by
\begin{eqnarray}
S^{co} & = & \exp \left\{ - \sigma_{co} \left[ N^{co}(b,s,y)-N^{sh}(b,s,y)\right] \, \ln \left[
  \frac{N^{co}}{N_{pp}(0)} \right] \right\} ,\nonumber \\
  N^{sh} & = &  \frac{\left( d \sigma^{c\bar{c}}_{pp} \big/ dy \right)^2}
{\sigma^{ND}_{pp} d \sigma^{J/\psi}_{pp} \big/ dy}  N_{bin}(b,s) S^{shad}(b,s,y)\,.
\label{eq:C}
\end{eqnarray}
Details of the model can be found in \cite{epjc08} and the
quantities in Eq.~(\ref{eq:C}) are all related to $pp$ collisions at
the corresponding energy and should be taken from experiment (the extrapolation for LHC is quite uncertain). The magnitude of recombination effect is controlled by the total charm cross section in $pp$ collisions and then dissociation-recombination effects will be of crucial importance in PbPb collisions at LHC.

At this stage is difficult to single out the size of each contribution at the LHC for the $J/\Psi \,(\Upsilon)+\gamma$. For such photon-associated high-$p_T$ quarkonium production, the final production cross section is modified by nuclear absorption, or final-state interaction between the pre-resonance state and the nuclear medium as discussed above. Notice that the Glauber absorption model is only valid for total production cross section which is dominated by low-$p_T$ quarkonium production. Here, we focus on the transverse momentum dependence of the nuclear modification factor due to gluon shadowing and it is timely to discuss the $p_T$ dependence of the final-state interaction. To address this issue, we can do an educated guess by verifying what we learn from RHIC. The modification factor $R^{J/\psi}_{AA}$ has been determined as a function of $p_T$ in the central rapidity region, analyzed both by the PHENIX \cite{PHENIX} and STAR \cite{STAR} collaborations. The factor increases with $p_T$ and STAR data indicate a nuclear modification
factor larger than one for $p_T \approx 8$ GeV \cite{STAR}. If this non-suppression of $J/\psi$ at large $p_T$ is confirmed it does not seem to behave as the other hadrons, which are significantly suppressed in central collisions and for increasing $p_T$. This behavior is closer to the one observed in prompt photons production. In the case of quarkonium plus a prompt photon, we then would expect the same non-suppresion pattern, since the hard photon is not affected by nuclear matter effects.

As a final comment, we notice that the present calculation can be also extended for the RHIC kinematic region, where the $q\bar{q}$ annihilation processes could also contribute. For instance, a study has  been done in Ref. \cite{Cooper} for the inclusive $J/\psi$ production in $pp$ collisions at RHIC within the PHENIX detector acceptance range using the color singlet and the color octet mechanisms based on NRQCD formalism. The calculations reproduce the RHIC data with respect to the $p_T$ distribution, the rapidity distribution and the total cross section at $\sqrt{s}=200$ GeV. In Ref. \cite{Elena}, the authors have considered the CSM approach (taking into account the $s$-channel cut contributions \cite{Lansberg_scut}) and cold nuclear matter effects to compute the $J/\psi$ nuclear modification factor for $dAu$ and $AuAu/CuCu$ collisions at RHIC energies. The data description is quite satisfactory. Therefore, this indicates the robustness of the framework in describing the high energy data. Of course, we are aware of the limitations of present calculation. The picture of a $J/\psi$  exactly recoiling back to back with the photon is simplified, since possible contributions of the color-octet
channel would lead to a fragmentation of the gluon into the meson, which would not be exactly recoiling against the photon. The experimental detection could be an additional concern.  For example, ALICE \cite{ALICE}  can tag the muons in the forward region, where
however there is no photon reconstruction, while it can detect the photons in
the central region, where however the muon detection is not as good. In the
case of CMS \cite{CMS}, the thresholds in $p_T$ both for muons and for the photon appear
quite low to ensure an efficient triggering.

In summary, in this letter we have investigated the quarkonium production in association with a isolated photon in pA and AA collisions at the LHC, considering the NRQCD factorization formalism and some of the parameterizations for the nuclear parton distributions available in the literature.  Our results show that the nuclear modification factors $R_{pA}$ and $R_{AA}$ are useful observables to discriminate among the different parton distributions in the nuclear medium. In particular, the predicted shadowing by the EPS parameterization is considerably larger than in the previous nuclear PDF's. The nuclear effects are different at central and at forward rapidities, so measuring the nuclear production of $J/\psi + \gamma$ and $\Upsilon + \gamma$ in these two regions could give additional insights about the correct nuclear gluon distribution. These complementary observables to the inclusive production are useful to shed light on quarkonium production where the lack of a good benchmark, both for proton-proton collisions, where the actual mechanism of quarkonium production is not known, and proton-nucleus collisions where the relative magnitude of the different effects (shadowing, nuclear absorption) cannot be fixed by present experimental data.
  
\begin{acknowledgments}
This work was partially financed by the Brazilian funding agency CNPq.
\end{acknowledgments}


\begin{thebibliography}{99}

\bibitem{CEM} J.F Amundson, O.J.P. $\acute{\rm{E}}$boli, E. M. Gregores and F. Halzen, Phys.\ Lett.\ {\bf B372} (1996) 127; G. A. Schuler and R. Vogt, Phys.\ Lett.\ {\bf B387} (1996) 181; C.~Brenner Mariotto, M.~B.~Gay Ducati and G.~Ingelman,
  Eur.\ Phys.\ J.\  C {\bf 23}, 527 (2002).

\bibitem{CSM} E.L. Berger and D.J. Jones. Phys. Rev. {\bf D23} (1981) 1521; R. Baier and R. Ruckl, Phys. Lett. {\bf B102}  (1981) 364.

\bibitem{NRQCD} G.T. Bodwin, E. Braaten, and G.P. Lepage, Phys.\ Rev.\ {\bf D51} (1995) 1125 [Eratum-ibid {\bf D55} (1997) 5853].

\bibitem{Lansberg} J.P. Lansberg, Int. J. Mod. Phys. {\bf A21} (2006) 3857; J.P. Lansberg, Eur.\ Phys.\ J.\  C {\bf 61}, 693 (2009).

\bibitem{Kramer:2001}
  M.~Kr\"amer,
  Prog.\ Part.\ Nucl.\ Phys.\  {\bf 47}, 141 (2001).


\bibitem{Kim1} M. Drees and C.S. Kim, Z. Phys. {\bf C53} (1992) 673.

\bibitem{Roy} D.P. Roy and K. Sridhar, Phys. Lett. {\bf B341} (1995) 413.

\bibitem{Basu} P. Mathews, K. Sridhar and R. Basu, Phys. Rev. {\bf D60} (1999) 014009.

\bibitem{Kim2} C.S. Kim, J. Lee and H.S.Song, Phys. Rev. {\bf D55} (1997) 5429.

\bibitem{Wang} R. Li and J.X. Wang, Phys. Lett. {\bf B672} (2009) 51.

\bibitem{Lansberg2} J.P. Lansberg, Phys. Lett. {\bf B679} (2009) 340.

\bibitem{vogt1}
 V.~Emel'yanov, A.~Khodinov, S.~R.~Klein and R.~Vogt,
  Phys.\ Rev.\ Lett.\  {\bf 81}, 1801 (1998).

\bibitem{vogt2}
   K.~J.~Eskola, V.~J.~Kolhinen and R.~Vogt,
  Nucl.\ Phys.\  A {\bf 696}, 729 (2001).

\bibitem{vogt3}
 S.~R.~Klein and R.~Vogt,
  Phys.\ Rev.\ Lett.\  {\bf 91}, 142301 (2003).


\bibitem{vogt4}
 R.~Vogt,
  Phys.\ Rev.\  C {\bf 71}, 054902 (2005).

\bibitem{vic_luiz1}
M.~B.~Gay Ducati, V.~P.~Goncalves and L.~F.~Mackedanz,
  Eur.\ Phys.\ J.\  C {\bf 34}, 229 (2004);  Phys.\ Lett.\  B {\bf 605}, 279 (2005).


\bibitem{ABMG:2006}
  A.~L.~Ayala Filho, C.~Brenner Mariotto and V.~P.~Goncalves,
  Int.\ J.\ Mod.\ Phys.\  E {\bf 16}, 1701 (2007).

\bibitem{BMG:2008}
  C.~Brenner Mariotto and V.~P.~Goncalves,
  Phys.\ Rev.\  C {\bf 78}, 037901 (2008).



\bibitem{arneodo}  M. Arneodo {\sl et al.},{ { Nucl. Phys}.} {\bf B483}, 3
(1997); {\ }{ Nucl. Phys.} {\bf B441}, 12 (1995).

\bibitem{e665}  M. R. Adams {\sl et al.}, {\ }{Z. Phys.} {\bf C67}, 403
(1995).

\bibitem{dglap}
 V. N. Gribov and L.N. Lipatov. { Sov. J. Nucl. Phys} {\bf 15}, 438 (1972);
 Yu. L. Dokshitzer. { Sov. Phys. JETP} {\bf 46}, 641 (1977);
 G. Altarelli and G. Parisi. { Nucl. Phys.} {\bf B126}, 298 (1977).


\bibitem{Mehen}
  T.~Mehen,
  Phys.\ Rev.\  D {\bf 55}, 4338 (1997)

\bibitem{Matrix} F. Maltoni {\it et al.}, Phys. Lett. {\bf B638} (2006) 202.


\bibitem{EKS98}
  K.~J.~Eskola, V.~J.~Kolhinen and P.~V.~Ruuskanen,
  Nucl.\ Phys.\  B {\bf 535}, 351 (1998);
  K.~J.~Eskola, V.~J.~Kolhinen and C.~A.~Salgado,
  Eur.\ Phys.\ J.\  C {\bf 9}, 61 (1999).



\bibitem{sassot}
  D.~de Florian and R.~Sassot,
  Phys.\ Rev.\  D {\bf 69}, 074028 (2004).



\bibitem{HKM}
  M.~Hirai, S.~Kumano and M.~Miyama,
  Phys.\ Rev.\  D {\bf 64}, 034003 (2001);
  Phys.\ Rev.\  C {\bf 70}, 044905 (2004).

\bibitem{HKN}
  M.~Hirai, S.~Kumano and S.~H.~Nagai,
  Phys.\ Rev.\  C {\bf 76}, 065207 (2007).


\bibitem{EPS08}
  K.~J.~Eskola, H.~Paukkunen and C.~A.~Salgado,
  JHEP {\bf 0807}, 102 (2008).



\bibitem{Braaten:2001}
  E.~Braaten, S.~Fleming and A.~K.~Leibovich,
  Phys.\ Rev.\  D {\bf 63}, 094006 (2001).

\bibitem{Elena} E.~G.~Ferreiro, F.~Fleuret, J.~P.~Lansberg and A.~Rakotozafindrabe, arXiv:0809.4684 [hep-ph].

\bibitem{Cap05} A. Capella and E. G. Ferreiro,  Eur. Phys. J. C {\bf 42}, 419 (2005).

\bibitem{epjc08}
A. Capella  {\it et al},   Eur. Phys. J. C {\bf 58}, 437 (2008);
K. Tywoniuk {\it et al.}, J. Phys. G: Nucl. Part. Phys. {\bf 35}, 104156 (2008).

\bibitem{PHENIX} M. Leitch [PHENIX Collaboration], Nucl. Phys. {\bf A830}, 27C (2009).

\bibitem{STAR} B. I. Abelev {\it et al.} [STAR Collaboration], Phys. Rev. {\bf C80}, 041902 (2009).

 \bibitem{Cooper} F. Cooper, M. X. Liu and G.C. Nayak, Phys. Rev. Lett. {\bf 93} (2004) 171801.

\bibitem{Lansberg_scut} H.~Haberzettl and J.~P.~Lansberg,
  Phys.\ Rev.\ Lett.\  {\bf 100}, 032006 (2008).

\bibitem{ALICE} B. Alessandro {\it et al}. [ALICE Collaboration], J. Phys. G: Nucl. Part. Phys. {\bf 32} (2006) 1295.

\bibitem{CMS} S. Chatrchyan {\it et al}. [CMS Collaboration], JINST {\bf 3}, S08004 (2008).






\end{thebibliography}
\end{document}